\documentclass[manuscript,screen]{acmart}

\AtBeginDocument{%
  \providecommand\BibTeX{{%
    \normalfont B\kern-0.5em{\scshape i\kern-0.25em b}\kern-0.8em\TeX}}}

\acmYear{2025}
\copyrightyear{2025}
\acmConference[ACM FAccT '25]{ACM Conference on Fairness, Accountability, and Transparency}{June 23--26, 2025}{Athens, Greece}
\acmBooktitle{ACM Conference on Fairness, Accountability, and Transparency (ACM FAccT '25), June 23--26, 2025, Athens, Greece}
\acmDOI{10.1145/3715275.3732003}
\acmISBN{979-8-4007-1482-5/25/06}

\usepackage{microtype}
\usepackage{geometry}
\usepackage{xcolor} 
\definecolor{main}{HTML}{5989cf}    
\definecolor{sub}{HTML}{cde4ff}     
\definecolor{white}{HTML}{ffffff}

\begin{document}

\title{Justified Evidence Collection for Argument-based AI Fairness Assurance}

\author{Alpay Sabuncuoglu}
\orcid{0000-0002-4415-0516}
\affiliation{%
  \institution{The Alan Turing Institute}
  \country{United Kingdom}}

\author{Christopher Burr}
\orcid{0000-0003-0386-8182}
\affiliation{%
  \institution{The Alan Turing Institute}
  \country{United Kingdom}}

\author{Carsten Maple}
\orcid{0000-0002-4715-212X}
\affiliation{%
  \institution{The Alan Turing Institute}
  \country{United Kingdom}}
\affiliation{%
  \institution{University of Warwick}
  \country{United Kingdom}}

\begin{abstract}
It is well recognised that ensuring fair AI systems is a complex sociotechnical challenge, which requires careful deliberation and continuous oversight across all stages of a system's lifecycle, from defining requirements to model deployment and deprovisioning. Dynamic argument-based assurance cases, which present structured arguments supported by evidence, have emerged as a systematic approach to evaluating and mitigating safety risks and hazards in AI-enabled system development and have also been extended to deal with broader normative goals such as fairness and explainability.
This paper introduces a systems-engineering-driven framework, supported by software tooling, to operationalise a dynamic approach to argument-based assurance in two stages. 
In the first stage, during the requirements planning phase, a multi-disciplinary and multi-stakeholder team define goals and claims to be established (and evidenced) by conducting a comprehensive fairness governance process. 
In the second stage, a continuous monitoring interface gathers evidence from existing artefacts (e.g. metrics from automated tests), such as model, data, and use case documentation, to support these arguments dynamically. 
The framework's effectiveness is demonstrated through an illustrative case study in finance, with a focus on supporting fairness-related arguments.
\end{abstract}

\ccsdesc[300]{Computing methodologies~Artificial intelligence}
\ccsdesc[100]{Software and its engineering~Software design techniques}
\ccsdesc[500]{Social and professional topics~Technology audits}

\keywords{trustworthy and ethical assurance, continuous fairness monitoring, system transparency artefacts, large language models in finance}



\maketitle

\section{Introduction}

AI-enabled software systems are applications or processes that integrate AI techniques, such as convolutional neural networks or large language models (LLMs), to enhance their functionality, decision-making capabilities, or efficiency \cite{amershi_guidelines_2019}. The term \textit{enabled} highlights that AI is typically one component within a larger, complex structure, providing learning capabilities such as data-driven pattern recognition or text generation. For instance, an AI-enabled financial market analysis system might use LLMs to process and summarise vast amounts of news articles, identifying sentiment patterns for specific organizations to incorporate as a "sentiment factor" within a broader, more complex analysis framework.

The increasing investment and demand for AI-enabled systems raises critical concerns about fairness, accountability, and transparency in the design, development, deployment and use of these systems \cite{bletchley_2023}. As a core part of our moral faculties \cite{haidt2007}, fairness underpins our perceptions of the trustworthiness of AI-enabled systems. Therefore, it is cruical that we ensure AI systems operate in a manner that aligns with societal values and norms. However, defining and operationalising fairness in AI-enabled systems is a complex and multifaceted challenge that requires careful consideration of technical, ethical, and social factors. 

In practice, the development and deployment of AI-enabled systems within organisations involve a complex network of roles, processes, and operational stages. These include upstream activities such as problem scoping and data collection (often involving domain experts, data engineers, and policy stakeholders), as well as downstream tasks such as model training, evaluation, integration, deployment, and monitoring (led by ML engineers, developers, product managers, and compliance officers). Fair AI is not just a technical challenge but also a societal imperative, requiring multifaceted and multistakeholder approaches to address the diverse and complex factors contributing to bias and discrimination. Recent studies build on earlier work to demonstrate the ongoing gap between operationalising fairness research into practical implementation, where practitioners struggle with balancing fairness-accuracy tradeoffs, resource limitations, and navigating complex sociotechnical contexts \cite{holstein_improving_2019, xivuri_how_2023}. 

Researchers and practitioners have proposed various frameworks, metrics, and interventions to address fairness, such as bias assessment and mitigation techniques, but gaps remain. Key challenges include the lack of standardized practices for data documentation (cf. model cards for model reporting), limited stakeholder engagement, and inadequate tools (or requisite skills and training) for continuous fairness monitoring. Developers of AI-enabled systems often face systemic challenges such as balancing the fairness-accuracy tradeoff \cite{wang2021}, optimising the quantitative notions of fairness \cite{verma_fairness_2018}, considering multiple sociotechnical aspects such as situational, ethical, and sociocultural context and policy \cite{selbst_fairness_2019}, and bridging the skill and research-capacity gap \cite{the_alan_turing_institute_2024_11092677}. These shortcomings highlight the need for robust and transparent mechanisms to assess and address fairness throughout the lifecycle of AI technologies.

In this paper, we propose a comprehensive (but modular) fairness review and monitoring approach that combines an argument-based assurance framework with dynamic evidence gathering using transparency artefacts. The evidence collection is facilitated by a software tool, which we made open-source on GitHub (\href{https://github.com/alan-turing-institute/fairness-monitoring}{alan-turing-institute/fairness-monitoring }). This approach is designed to integrate into an organisation’s AI system development lifecycle, supporting decision-making from early design through deployment and ongoing monitoring.  It aims to address fairness challenges in AI systems by systematically defining trustworthiness goals, collecting evidence to support these goals, and monitoring the system’s performance over time. The tools are also designed for seamless integration into existing organisational workflows, including risk management structures. We demonstrate the applicability of this approach through heuristic walkthroughs of case studies in finance and healthcare, highlighting the importance of transparency, accountability, and stakeholder engagement in the development process. We discuss how the framework can be adapted for different organisational contexts and stages of maturity.

\section{Background}

At its core, fairness in AI seeks to advance two goals that capture positive and negative moral duties. The negative duty is to ensure that systems operate without unacceptable forms of bias or discrimination against individuals or groups based on race, gender, socioeconomic status, or other protected attributes. While legal and regulatory frameworks define and rely on protected characteristics (e.g. EU AI Act) as a means for operationalising 'acceptable' versus 'unacceptable' forms of bias, this still involves complex normative judgments (e.g. where varying base rates between groups are known to be unequal and relevant to, say, a classification task) and must be contextualised within specific domains and use cases (e.g. healthcare, finance, criminal justice) \cite{gallegos_bias_2023, selbst_fairness_2019}. Transparency in how these trade-offs and judgements (or assumptions) have been handled, therefore, continues to be a good approach for helping build ethical capabilities within the assurance ecosystem \cite{habli_25}. The positive moral duty seeks to ensure AI also advances goals such as social justice and equity, rather than merely avoiding harms. However, achieving fairness in either direction is complex because of the myriad risks and opportunities that exist across a project's lifecycle and the global reach of some AI systems. 

This latter issue intersects with ongoing societal debates around what constitutes "fairness" reveal conflicting norms or priorities for different groups and communities, such as individual merit versus group equity, making consensus difficult \cite{speicher_unified_2018}. Multiple studies across time have confirmed the challenge of achieving consensus on fairness definitions, with recent research on LLMs showing these tensions persist in emerging technologies and show little sign of abating \cite{speicher_unified_2018, gallegos_bias_2023, xivuri_how_2023}.Translating existing societal challenges into algorithmic systems cause some common issues such as failing to capture the depth of the problem. Selbst et al. coins this term as \textit{portability trap}, highlighting the dangers of transferring algorithmic solutions between differing social contexts where underlying assumptions may not hold \cite{selbst_fairness_2019}. When AI systems are deployed in such environments, they risk perpetuating these inequities, as they rely on existing technological infrastructure, historical data, and other inherited technical interventions that may be inherently biased. 

The history of moral and political philosophy, as well as domains such as jurisprudence, should be sufficient grounds to convince the skeptic that there is no easy technological fix to the problem of AI fairness. However, we can also point to specific issues such as how improving fairness can lead to reductions in overall accuracy, creating tension for organizations prioritizing efficiency or profitability \cite{xivuri_how_2023}. Or, additionally, how fairness interventions can be computationally expensive and require significant expertise, limiting their adoption \cite{holstein_improving_2019}. The result is that AI developers need to choose between fairness notions to evaluate their model performance with careful consideration given to any trade-offs, and to then justify why this decision was acceptable within the intended context of use. Empirical research with AI practitioners has consistently demonstrated that successful fairness implementation requires collaborative approaches involving diverse stakeholders throughout the development lifecycle, whether because of skills and knowledge gaps or other competing pressures (e.g. time to product deployment) \cite{holstein_improving_2019, xivuri_how_2023}.

Moreover, while the unfairness of algorithmic systems, particularly AI-enabled systems, has received significant attention, a notable gap persists between research advancements and practical implementation \cite{veale_fairness_2018, holstein_improving_2019}. A commonly recognised issue is the disconnect between the real-world obstacles faced during the development process and the assumptions underlying much of the fair ML literature. A holistic and proactive approach to tackling fairness issues with detailed granularity while maintaining a system-level perspective is much needed \cite{holstein_improving_2019}. Fairness in AI cannot be achieved in isolation but must involve active collaboration among technical experts, domain specialists, policymakers, and affected communities. Establishing a space where all stakeholders can freely express their views, challenge assumptions, and contribute to a shared understanding of the AI system's goals and potential impacts is a key requirement throughout the fair AI development process \cite{xivuri_how_2023}.

This paper introduces a proactive fairness monitoring approach that leverages argument-based assurance techniques and existing metadata as justified evidence to enable structured and transparent sharing of decisions and outcomes.

\section{\textit{Argument}-based Assurance with Proactive Evidence Collection}

Argument-based assurance is a methodology that aims to demonstrate the rationale behind decisions or actions taken throughout the lifecycle of a system, using a \emph{structured argumentation} process to document and explain how claims made about a system's goals are justified by evidence. The approach has been used for several decades in the context of safety-critical engineering (e.g. manufacturing and aviation)\cite{civilaviationauthority2010}. More recently, in the domain of AI safety, this approach has been popularised by several organisations including governmental departments \cite{clymer_safety_2024, buhl_safety_2024}. 
This process results in the creation of an assurance case, a document that sets out the argument for the system's trustworthiness. 

However, while the traditional focus has been on normative goals such as safety or security, several research groups have started exploring how the same methodology can be used to ensure a systematic approach to broader ethical goals such as fairness or explainability \cite{burr2021ethical, porter2024principles}. Key to these efforts is a focus on supporting multi-stakeholder engagement (e.g. helping evaluate the sufficiency of key design requirements) and internal deliberation (e.g. identifying gaps in a provisional argument prior to formal evaluation and deployment). The purpose of this is twofold: (1) To help promote an inclusive approach to the operationalisation of normative goals such as `fairness', by promoting a broader diversity of considerations to be reflected in the scope of an argument. (2) To shift the focus of assurance cases away from a form of compliance towards a more reflective approach to responsible research and innovation.

To support these goals, and building on prior work, we have developed an interactive platform, known as the Trustworthy and Ethical Assurance (TEA) platform which helps teams with limited experience and resources to follow a structured approach from the early stages of the development process. Before we discuss how this platform can be extended to accommodate dynamic forms of AI fairness assurance, it's worth briefly explaining the core components of an assurance case for those who are unfamiliar with the methodology. 

\subsection{Technical Implementation}

The interactive TEA platform is hosted on Azure, utilising Django and Postgres for backend and Next.js for frontend development. Open-source code and documentation is also available through GitHub (\href{https://github.com/alan-turing-institute/AssurancePlatform}{alan-turing-institute/AssurancePlatform}) for alternative deployments. The platform is accompanied by openly accessible tutorials and curricula to help widen user engagement. The evidence collection framework has minimal dependencies (pandas, PyYAML, Jinja2) to ensure seamless integration with the existing workflows (Github: \href{https://github.com/alan-turing-institute/fairness-monitoring}{alan-turing-institute/fairness-monitoring}).

\begin{figure*}[h!]
\centering
\includegraphics[width=.75\linewidth]{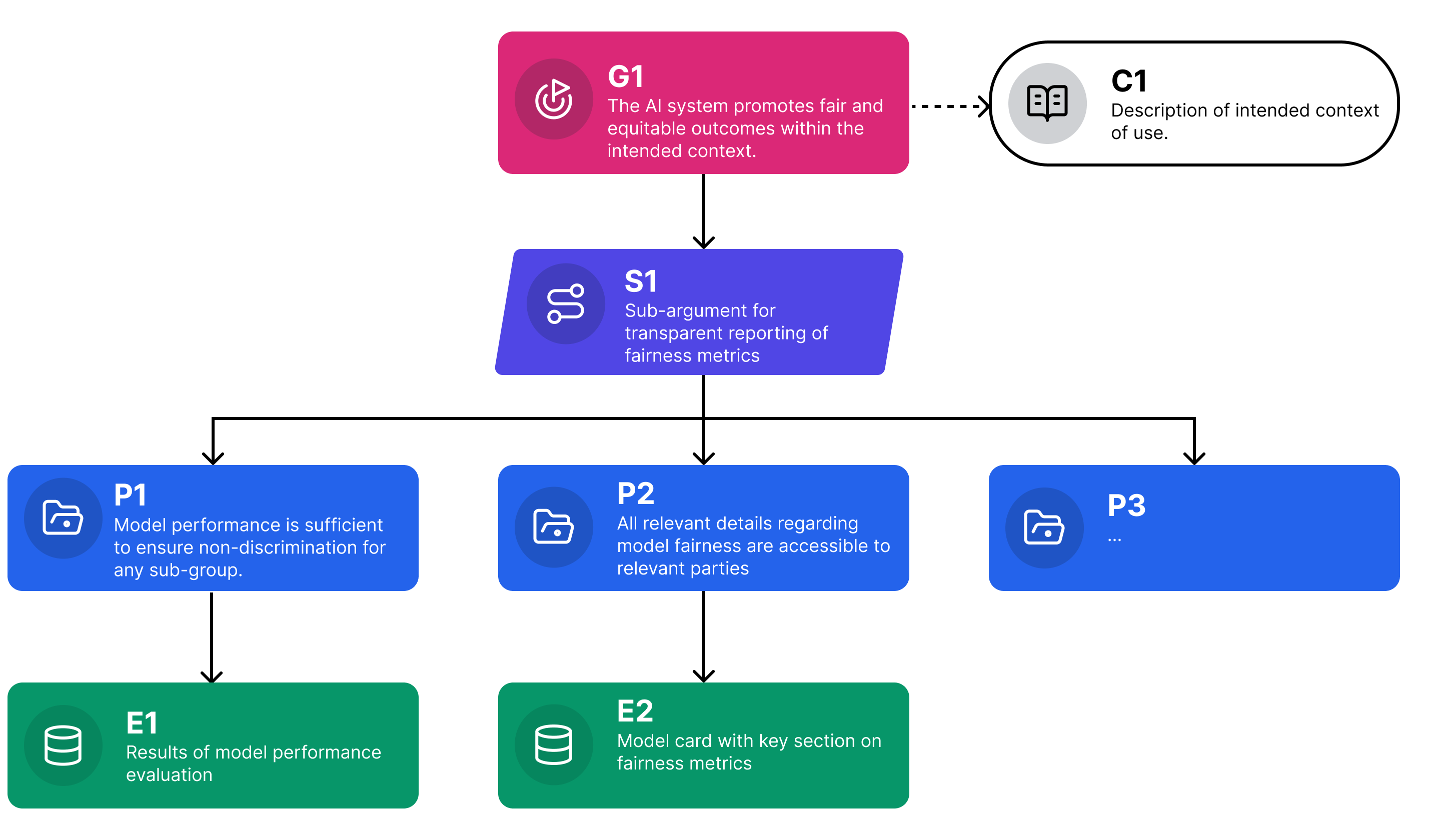}
\caption{\label{fig:acexample}A toy model showing an incomplete example of an assurance case for a fairness-related goal.}
\end{figure*}

\subsection{Core Components of an Assurance Case}

The main components of an assurance case with some practical guidance for development teams are:

\subsubsection{Goal Claim:} This claim directs the focus of the assurance case towards a desired value or principle of the system. For example, a goal claim could be that an AI system is ``fair'' or ``explainable''. The goal claim should be specific and clearly defined to avoid ambiguity---although full specification and operationalisation of the term emerges in the subsequent argument.

\noindent \fcolorbox{main}{sub}{ 
\begin{minipage}{\dimexpr\linewidth-2\fboxsep-2\fboxrule} 
    Define goals around trustworthiness characteristics and relevant legal or regulatory requirements. For example, a development team in the EU could start with frameworks like NIST's AI RMF \cite{tabassi_artificial_2023} and the EU AI Act \cite{eu_2024}. 
\end{minipage}
}

\subsubsection{Context:} Context elements define the specific conditions under which the claims are being made. For example, the context of an AI system used by healthcare professionals in a hospital would be different from an AI system used by patients at home. Making the context explicit helps in evaluating the validity of the claims and the sufficiency of the evidence, because no goal can be assured universally (e.g. a plane is not `safe' to fly by untrained civilians; an AI system may not be `explainable' by or to all people).

\noindent \fcolorbox{main}{sub}{ 
\begin{minipage}{\dimexpr\linewidth-2\fboxsep-2\fboxrule} 
    Collaborate with business, design, and user research teams to structure the context. Insights from user researchers can guide considerations for AI-enabled versus rule-based solutions. 
\end{minipage}
}

\subsubsection{Property Claims:} These are lower-level claims that specify the properties of the system or project that contribute to achieving the goal claim. For example, property claims for an "explainable" AI system could include the techniques used to interpret the model's outcomes or whether users can contest decisions based on the explanations provided. 

\begin{figure*}
    \centering
    \includegraphics[width=\linewidth]{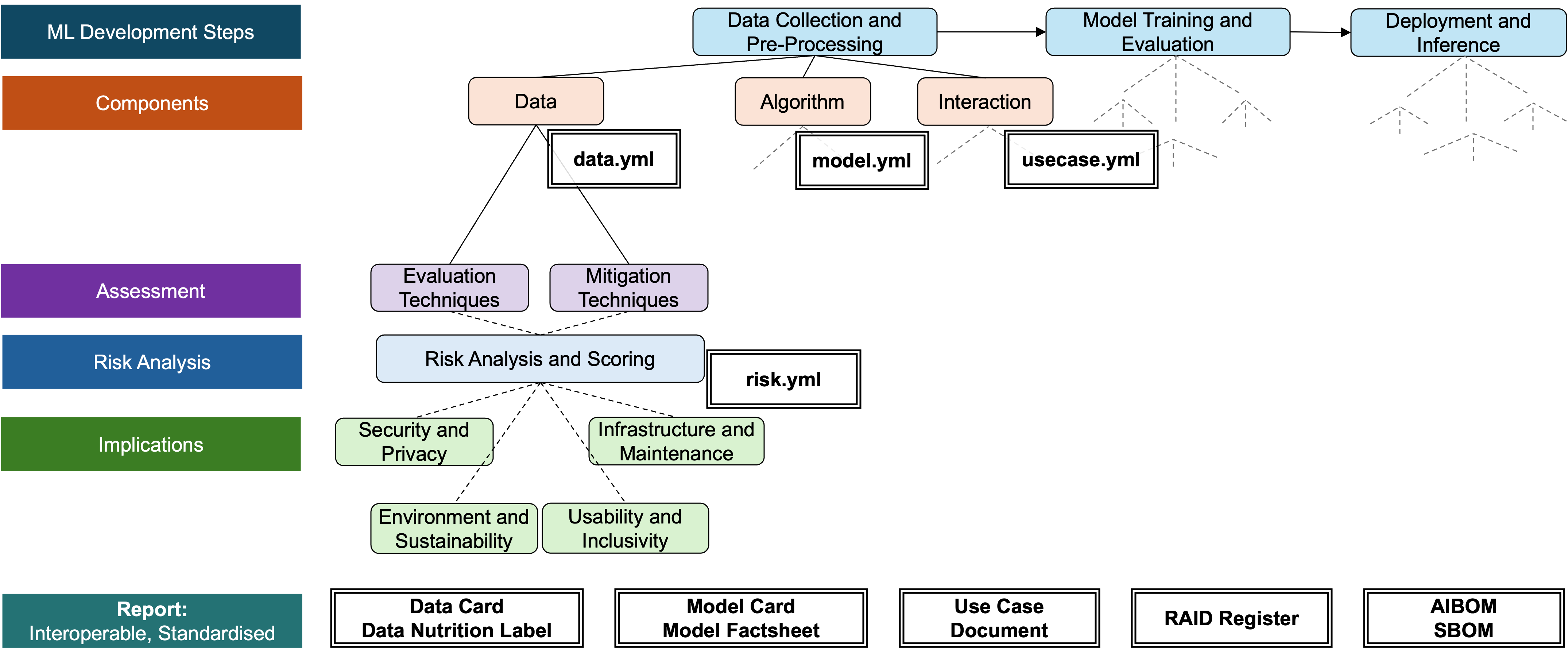}
    \caption{From structured arguments to collecting evidence: In the argument-based assurance process, we defined property claims using a hierarchical mechanism based on ML development stages and components. This figure illustrates this hierarchical process and points respective transparency artefacts for evidence collection.}
    \label{fig:structured}
\end{figure*}

\noindent
\fcolorbox{main}{sub}{
\begin{minipage}{\dimexpr\linewidth-2\fboxsep-2\fboxrule} 
    \label{box:property}
    Conduct a thorough literature review and adapt existing guidelines to the specific use case. Developing property arguments is often the most challenging step. It requires discussions among developers and stakeholders about potential issues and mitigation strategies. To streamline this process, we propose a hierarchical mechanism comprising four levels: (1) ML Stages, (2) Components, (3) Assessment, and (4) Implications. Figure \ref{fig:structured} illustrates this structured process with its relation to Evidence collection. 
\end{minipage}
}

\subsubsection{Strategy:} This element outlines the reasoning and approach used to develop the argument supporting the goal claim. Strategies don't contain any substantive content on their own, but help break down the argument into related sub-arguments, making it easier to understand and evaluate. 

\noindent
\fcolorbox{main}{sub}{
\begin{minipage}{\dimexpr\linewidth-2\fboxsep-2\fboxrule} 
    Formulate strategies in tandem with property arguments, akin to defining abstract objects in software engineering.
\end{minipage}
}

\subsubsection{Evidence:} This is the foundation of the assurance case and provides the basis for trusting the validity of the claims. Evidence can include empirical data, expert opinions, or adherence to technical standards. The type of evidence required will depend on the specific claims being made.

\noindent
\fcolorbox{main}{sub}{
\begin{minipage}{\dimexpr\linewidth-2\fboxsep-2\fboxrule} 
    Determine the types of evidence required and establish efficient and continuous collection methods. Automating this process can help define pass-fail tests integrated into the model development lifecycle. 
\end{minipage}
}

Structured arguments can support development teams address the portability trap in fairness \cite{selbst_fairness_2019} by making assumptions explicit. Figure \ref{fig:acexample} shows an (incomplete) toy model of how these core components could be used to begin developing a structured argument for justifying and operationalising a high-level goal, such as ``The AI system promotes fair and equitable outcomes within the intended context.'' Additional elements would need to be added, including context statements that help define the AI system itself, additional strategies (e.g. to provide shape to the argument and ensure a key focus on claims related to, say, multi-stakeholder engagement or participatory design, are include), and of course further claims (e.g. P3) that would help operationalise and specify the notion of fairness that is appealed to in the goal claim (G1).

In this argument-based assurance process, types of evidences might vary in terms of relevance, completeness, admissibility, and accuracy \cite{ashmore_assuring_2022}. Formal verification techniques can demonstrate a full coverage and guarantee that the fairness argument is valid for the current state of the model \cite{borca-tasciuc_provable_2022,yadav_fairproof_2024}. These techniques can also analyse the model without requiring access to the training or evaluation data, which is beneficial when data privacy is a concern. However, these methods are also hard to scale as they involve complex geometric algorithms in high-dimensional spaces.

An alternative approach is providing justified evidence which refers to documented, verifiable information that supports arguments about an AI system's performance, safety, fairness, and trustworthiness \cite{DSIT_assurance}. In this case, the evidence must be based on measurable and reproducible outcomes, such as evaluation metrics, benchmarking results, or test data. It ensures stakeholders can rely on the system's integrity and effectiveness by providing clear, valid proof for key assurances such as accuracy, robustness, and ethical compliance.

\subsection{Collecting Evidence}

Throughout the development of ML models, developers and other stakeholders use various documentation formats to enhance reproducibility and communicate the details of artefacts with both internal and external stakeholders. Organizations use metadata recording formats, such as model cards, data cards, and algorithmic transparency frameworks to improve transparency across development and deployment workflows. We refer to these documentation tools as "transparency artefacts," which are designed to enhance clarity, accountability, and trust in ML systems.

We can use these transparency artefacts as justified evidence to verify the team took the required actions and created enough evidence for the given arguments. The justified keyword is a reference to a recent concept of "justified trust" \cite{DSIT_assurance}, referring a level of trust that AI system's safety and compliance is based on clearly communicating reliable evidences to stakeholders. The key requirement for this kind of evidence is they must be grounded in measurable and reproducible outcomes such as benchmarking results, auditing trails, or red teaming reports. We provide a use case in Section \ref{sec:finsentiment}.

\subsubsection{Data Cards}

Data cards provide detailed documentation of datasets used in training and evaluating ML models. They include information about data sources, preprocessing steps, limitations, and known biases, offering critical transparency about the data shaping model behaviour. This transparency is a valuable resource for identifying potential biases, evaluating model fairness, and understanding the broader sociotechnical challenges associated with the dataset \cite{pushkarna_data_2022}.

\noindent \fcolorbox{main}{sub}{ 
\begin{minipage}{\dimexpr\linewidth-2\fboxsep-2\fboxrule} 
\textbf{Fairness-related metadata:} Google's Data Card specification mandates details such as funding sources, data subjects, representation balances of potential sensitive characteristics, dataset collection process, geographies involved in both collection and labelling processes, intended use cases, and potential unintended outcomes.
\end{minipage}
}

\subsubsection{Model Cards}

Model cards provide essential information about an ML model's intended use, limitations, and performance. They typically include details like model purpose, dataset used, evaluation metrics, ethical considerations, and any limitations or potential biases. To illustrate, Google's model card specification\footnote{\url{https://modelcards.withgoogle.com/about}} includes nine main sections \cite{mitchell_model_2019}: \textbf{(1)} Model details such as basic model information, \textbf{(2)} Intended use with the cases that were envisioned during development, \textbf{(3)} Factors could include demographic or phenotypic groups, environmental conditions, technical attributes, \textbf{(4)} Metrics chosen to reflect potential real world impacts of the model, \textbf{(5)}Evaluation data that was used for the quantitative analyses in the card, \textbf{(6)} Training data (when possible) to understand the distribution over various factors, \textbf{(7)} Quantitative analyses, \textbf{(8)} Ethical considerations, and \textbf{(9)} Caveats and recommendations. These information help users and other stakeholders to understand where and how a model should (and shouldn't) be applied.

\noindent \fcolorbox{main}{sub}{ 
\begin{minipage}{\dimexpr\linewidth-2\fboxsep-2\fboxrule} 
\textbf{Fairness-related metadata:} ``Data``, ``Performance Metrics``, and ``Considerations`` field groups can be a valuable source to identify potential fairness issues. Entities with ``Data`` field can reveal potential representation bias and understand the sensitive characteristics. ``Performance metrics of quantitative analysis`` can include sensitive-group based analysis. ``Confidence intervals`` show the statistical significance of these results. ``Considerations`` can support the next phase of requirement planning in an iterative development process.  ``License information`` and its ``SPDX`` (System Package Data Exchange) link can reveal SBOM (software bill of materials) data for their development codebase. An SBOM document lists all the components that forms a software, including: Dependencies like libraries and API calls, and their versions and licenses. Particular versions of development libraries can introduce some measurement and evaluation bias throughout the development process.
\end{minipage}
}

\begin{figure*}
    \centering
    \includegraphics[width=.8\linewidth]{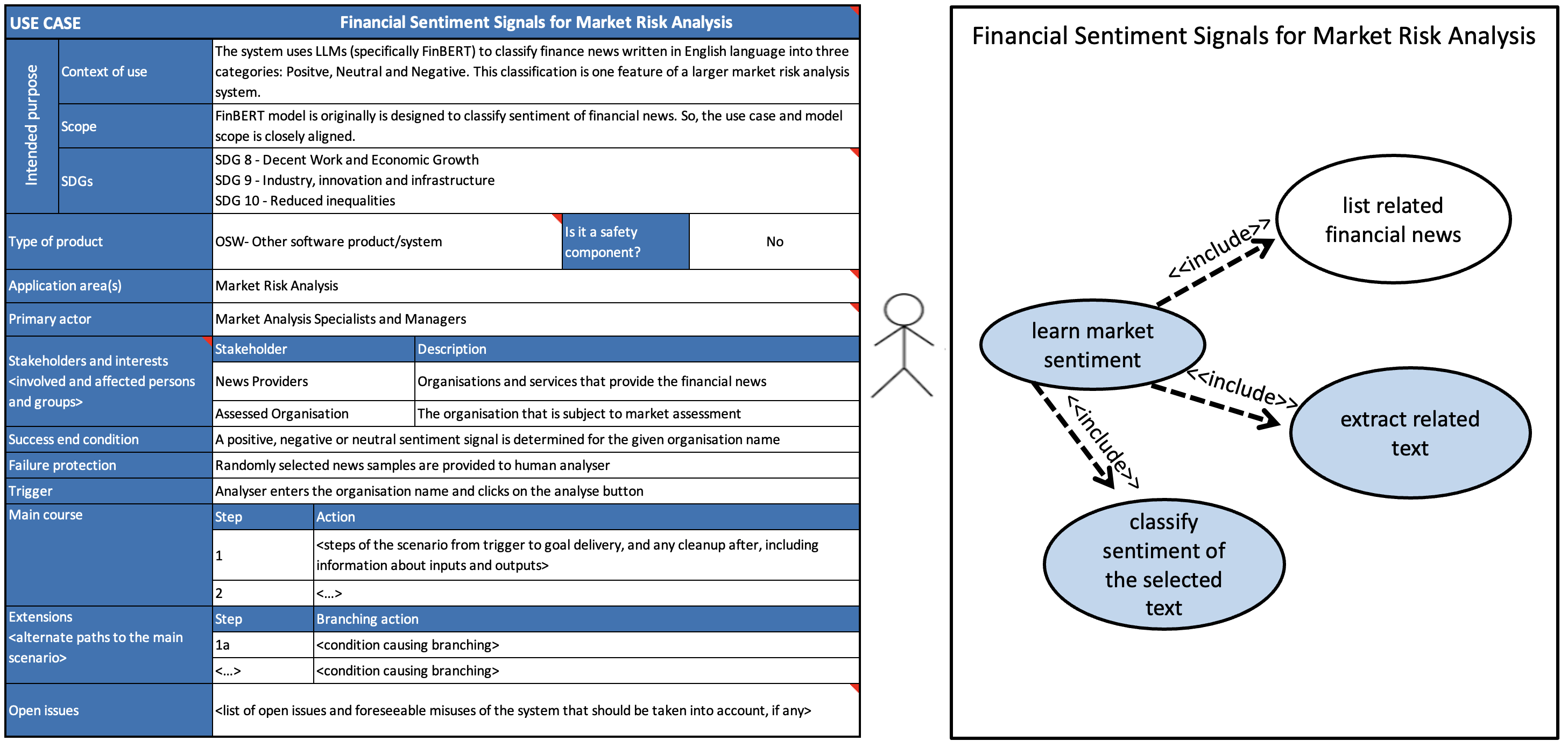}
    \caption{A simplified example use case card for a financial sentiment analysis system. The template is obtained from \cite{hupont_use_2023}.}
    \label{fig:usecasecard}
\end{figure*}

\subsubsection{Use Case Cards}

Based on the commonly used Unified Modelling Language (UML) in software engineering, Hupont et al. proposed use case cards as a standardised methodology to define intended purposes and operational uses of an AI system \cite{hupont_use_2023}. A use case card consists of two main parts: (a) a canvas for visual modeling, and (b) a table for written descriptions. The canvas includes actors, use cases, and relationships, while the table provides additional details such as intended purpose, product type, safety component status, application areas, and other relevant information. Figure \ref{fig:usecasecard} illustrates a use case card created for a financial sentiment analysis system, that we conducted a heuristic walkthrough in Section \ref{sec:finsentiment}

\noindent \fcolorbox{main}{sub}{ 
\begin{minipage}{\dimexpr\linewidth-2\fboxsep-2\fboxrule} 
\textbf{Fairness-related metadata:} The EU AI Act alignment of the proposed use case card, and direct links to open issues gives a valuable system and organisation level view on fairness challenges. 
\end{minipage}
}

With our approach to argument-based assurance and several examples of evidence illustrated introduced, we now turn to an illustrative example of how the methodology and transparency artefacts can be consolidated.

\subsection{Recording Fairness-Related Experiment Metadata}

Although, model, data and use case cards store some fairness-related metadata, they are not designed to address potential fairness recording needs. We created a fairness recording template to support development teams documenting key details related to the experimental setup, data characteristics, model specifications, and fairness evaluation metrics. The metadata begins with general information, providing a comprehensive overview of the experimental context.

The "data" section captures the characteristics of the dataset, such as "sample" details and profiles of key "variables". It explicitly identifies "sensitive\_characteristics" (e.g., race, gender) that may influence fairness outcomes, alongside the categorization of "nominal" and "continuous" features. The "model" section records the ML model's "name", ensuring traceability to the specific algorithm or architecture used. In the "sample\_data" subsection, performance metrics such as true positives ("tps"), false positives ("fps"), true negatives ("tns"), and false negatives ("fns") are recorded, providing a foundation for quantitative analysis of model behaviour.

Finally, the "bias\_metrics" section is pivotal in evaluating fairness. It organizes metrics by "groups", where each "group\_name" corresponds to a demographic or attribute category (e.g., age group, gender). Within each group, metrics are detailed with attributes such as their "name", "description", "value", and corresponding "thresholds". Parameters such as whether a higher metric value is preferred ("bigger\_is\_better") and additional "notes" or "subgroup (sg)" parameters further enrich the analysis. By facilitating granular tracking and assessment of fairness across groups, this template provides a robust mechanism for identifying and addressing bias, ultimately fostering ethical and responsible ML development.

\section{Heuristic Walkthrough: AI-Enabled Financial News Sentiment Analysis}
\label{sec:finsentiment}

The use case is a heuristic walkthrough of our two-step methodology, showcasing its application in the finance sector. A heuristic walkthrough is an evaluation method that involves a step-by-step analysis of a system or process using predefined heuristics to identify potential usability (and other characteristics of interest) issues. \cite{sears1997heuristic}. This paper uses this technique to a) evaluate the assurance methodology by outlining the key principles as heuristics, b) conduct the walkthrough to evaluate its application on different scenarios, and c) document findings.

\begin{figure*}[h!]
    \centering
    \includegraphics[width=.75\linewidth]{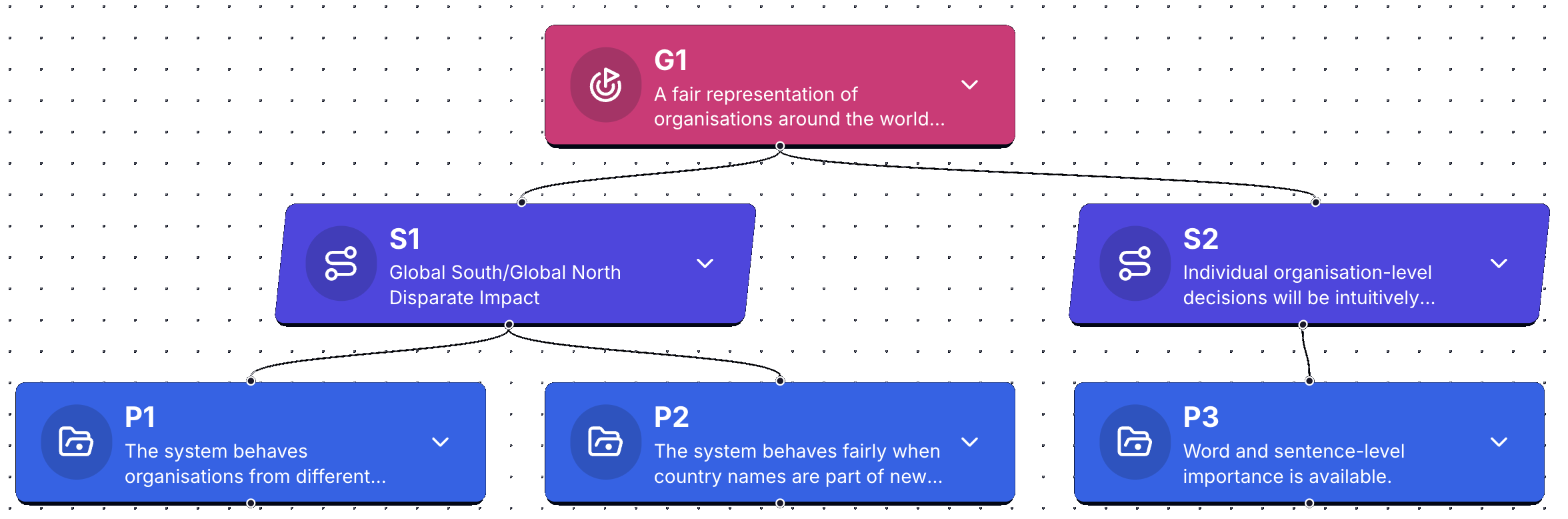}
    \caption{Creating arguments to achieve a fairer financial sentiment analysis system.}
    \label{fig:tea-finance}
    \vspace{-0.5cm}
\end{figure*}

\subsection{Use Case Context and Challenge}

In financial services, sentiment signals are primarily used to inform market investment decisions \cite{10371970}. While biases in these signals may not be immediately apparent, they can have significant long-term consequences for individuals when scaled across markets or entire economies. A recent report on the impact of LLMs in financial services highlights that achieving fairness requires both technical expertise and social awareness within development teams \cite{maple_impact_2024}. In the development of LLMs and other ML models, researchers and developers focus on defining fairness in social and technical terms and measuring it quantitatively. However, when it comes to LLMs, many institutions are still exploring whether assessing fairness in natural language outputs differs from other types of models and if this task is inherently more complex \cite{gallegos_bias_2023}.

LLMs have been utilised by finance organisations and public bodies to improve the news sentiment analysis process, where models analyse vast quantities of financial and political news to generate sentiment signals that inform investment and risk management decisions \cite{maple_impact_2024}. LLMs can benefit the overall performance by better understanding complex language and interpreting quantitative data, economic indicators, and historical trends. However, the challenge here lies in ensuring the model interprets sentiment fairly and accurately across multiple contexts.

\subsection{Defining Goal Claims}

When analysing data sources from different geographies, such as financial news and company finance reports, the model may risk reflecting biases inherent in its language, technical jargon, or structure. These implicit bias sources can lead to more positive interpretation of data dominant regions, in this case the Global North, potentially underrepresenting views or concerns relevant to countries in the Global South. Global South/ North divide is one definition to characterise the sensitive groups in financial sentiment analysis. It is also possible to create different groups based on access to socioeconomic benefits, or inclusion to news sources. In this use case, we define this global south/north binary group based on UN Trade and Development definition \cite{unctad_forging_2018}. 

\subsection{Defining Strategies}

In this use case, a misguided investment decision can result in an organisation or country level impacts. The fairness impact of this kind of decision is not so visible instantly. A biased decision can result in lack of investment, and as a result, the direct individual impact is generally potential loss of financial wellbeing. 

One way to define a strategy is using existing fairness notions as different strategies to group our data, model, interaction based property claims. In this use case, we utilised Franklin et al.'s fairness metric ontology \cite{fontology} to define our strategies defined around group and individual fairness definitions. Using this ontology also allowed us to link the right metrics while collecting evidences for the given property claims and strategies.

In the end, considering our goal and definition around GS/GN divide, we created our properties around group and invidual fairness strategies, as seen in Figure \ref{fig:tea-finance}.

\subsection{Defining Property Claims}

As we elaborated in Section \ref{box:property}, we structured our property claims around data, model, and interaction components. These claims do not cover all stages of the system development and only limited to our use case tasks:

\begin{itemize}
  \item \textbf{Data Component Arguments:} Data involved in developing AI models can be various including training, testing, validation, fine-tuning and reference data. Data component arguments focus on ensuring that the data used for developing and operationalising the sentiment analysis model is representative and unbiased.
  \begin{enumerate}
    \item \textit{Global South Representation}: The data includes a diverse range of financial news sources from both the Global South and North to ensure balanced sentiment analysis across regions.
    \item \textit{Diverse Perspectives}: The model is trained on a dataset that includes a variety of perspectives and viewpoints, ensuring that sentiment signals are not skewed towards specific regions or sources.
  \end{enumerate}
  \item \textbf{Model Component Arguments:} The model component arguments focus on ensuring that the sentiment analysis model is accurate, interpretable, and fair across different regions and contexts.
  \begin{enumerate}
    \item \textit{Interpretability}: The model provides clear, interpretable sentiment signals that can be easily understood and validated by financial analysts and regulators.
    \item \textit{Global Fairness}: The model generates sentiment signals that are fair and accurate across different regions, ensuring that the analysis is not biased towards specific countries or regions.
    \item \textit{Confidence Scores}: The model provides confidence scores for sentiment signals, allowing users to assess the reliability and accuracy of the analysis.
  \end{enumerate}
  \item \textbf{Interaction Component Arguments:} The interaction component arguments focus on ensuring that the sentiment analysis model is transparent, accountable, and ethically deployed in financial decision-making processes.
  \begin{enumerate}
    \item \textit{Transparency Artefacts}: The model generates transparency artefacts, such as interpretability reports and confidence scores, to provide insights into the sentiment analysis process and ensure accountability.
    \item \textit{Regulatory Compliance}: The model complies with financial regulations and guidelines, ensuring that sentiment signals are generated in a manner that is consistent with legal and ethical standards.
    \item \textit{Stakeholder Engagement}: The model engages with financial analysts, regulators, and other stakeholders to gather feedback and insights on the sentiment analysis process, improving trust and accountability.
  \end{enumerate}
\end{itemize}

In this use case, we selected:

\begin{itemize}
    \item \textbf{FinBERT model}\footnote{\url{https://huggingface.co/yiyanghkust/finbert-tone}} \cite{huang_span_2023}, an LLM trained on financial text, to analyse the financial news data. We selected this model as it is a) popular and b) open-source, c) has transparent dataset documentation, and d) requires small compute-requirement for easy and sustainable experiment reproducibility.
    \item \textbf{Indian financial news dataset\footnote{\url{https://huggingface.co/datasets/kdave/Indian\_Financial\_News}}} to evaluate the model's performance on a data source produced in a different geography. The dataset includes equal number of samples (n=8987) from each sentiment tone category: "Positive", "Negative" and "Neutral". This evaluation against a real-word, different country-sourced dataset is important to understand the model's performance on different technical jargon, currency, and other implicit bias sources. 
\end{itemize}

\begin{figure*}
    \centering
    \includegraphics[width=\linewidth]{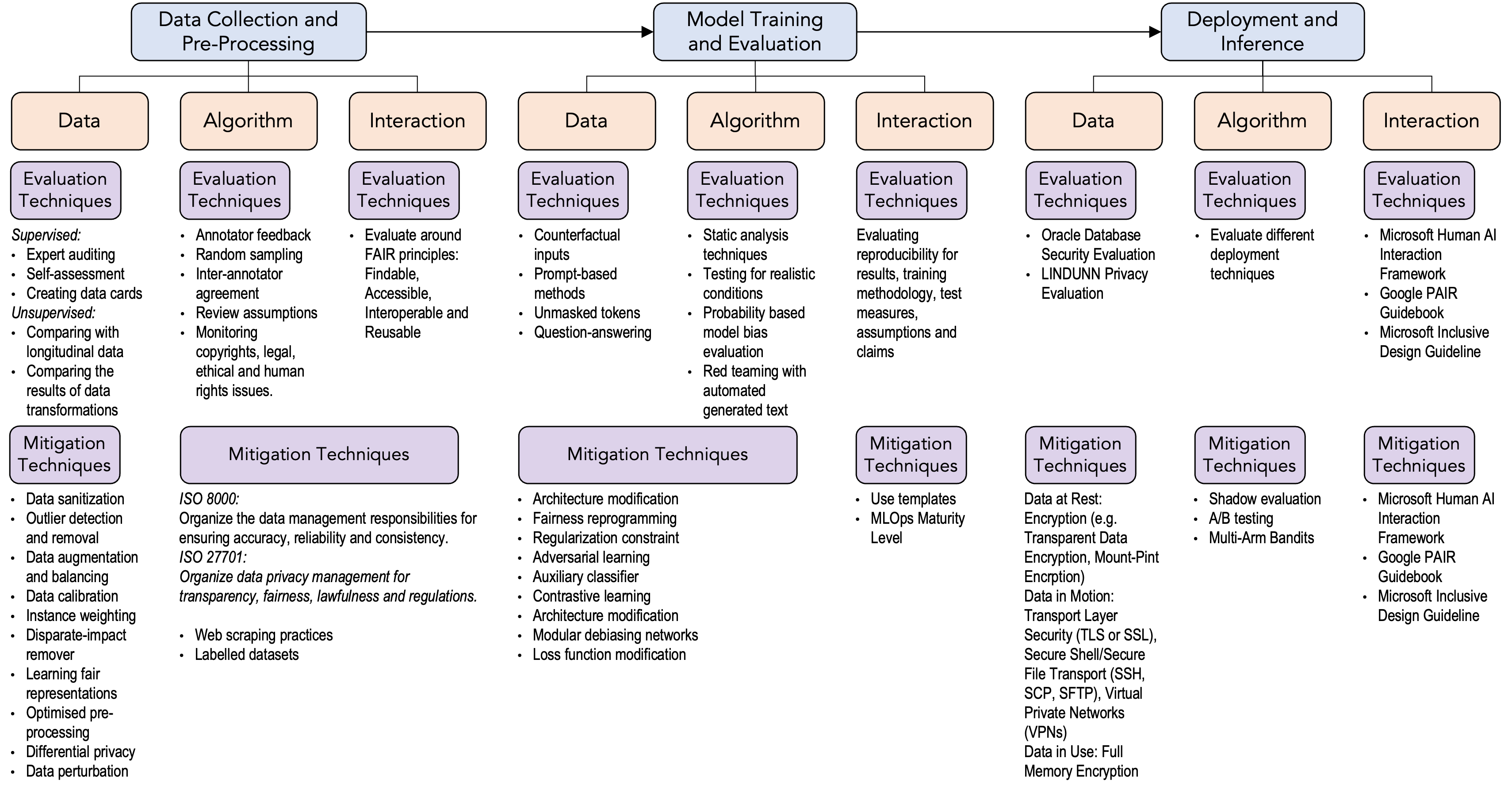}
    \caption{Categorical diagram of selected evaluation and mitigation techniques that can address fairness issues directly in a component-based risk analysis. }
    \label{fig:llmmethods}
\end{figure*}

\subsection{Collecting the Evidence}

Fairness evaluation is a continuous process that requires regular check-ups based on the latest regulation and sociotechnical issues. Figure \ref{fig:llmmethods} illustrates a list of potential evaluation experiments and mitigation actions that can be used during the risk assessment and mitigation process. The list is not exhaustive, and based on selected ISO standards (ISO 27001 and 24027), OWASP guidances (Top 10 for LLMs\footnote{\url{https://owasp.org/www-project-top-10-for-large-language-model-applications/}}), and Gallegos et al.'s review \cite{gallegos_bias_2023}. We share this list to help reader visualise the potential sources of evidences that can be linked to different components of different stages in ML development process.

\begin{figure*}[h!]
  \centering
  \includegraphics[width=\linewidth]{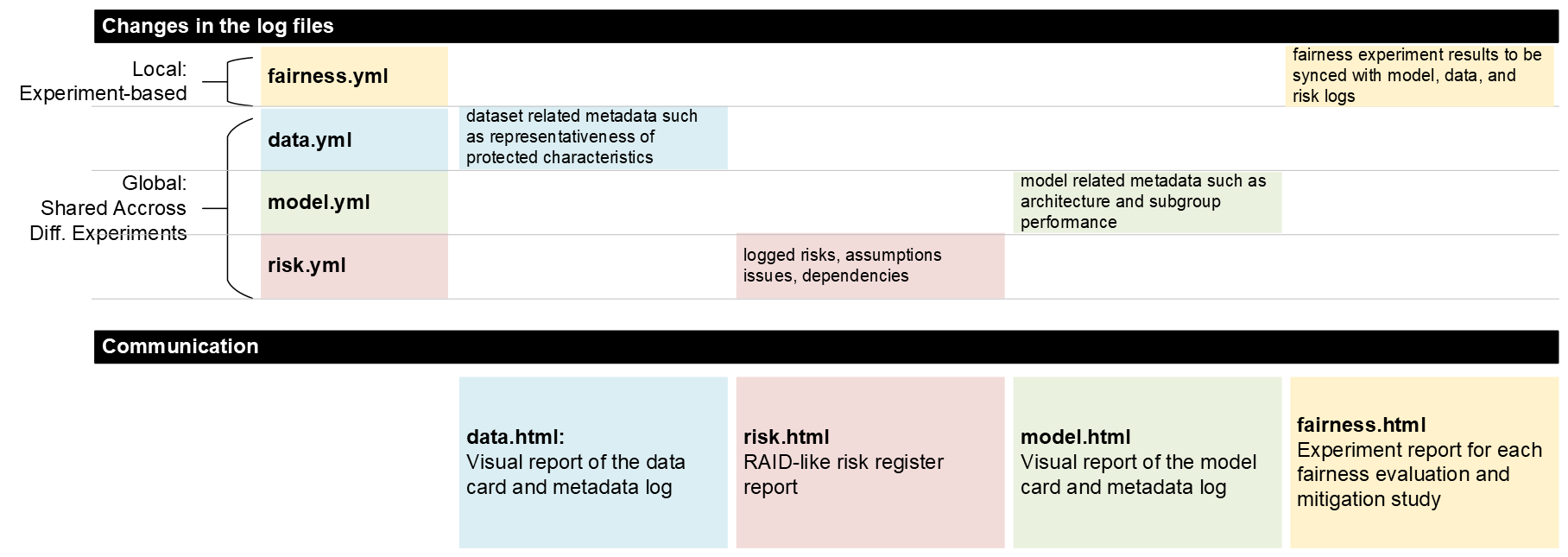}
  \caption{The changes in the log file and generated reports throughout the project timeline.}
  \label{fig:metadata}
\end{figure*}

We selected three tasks that most development teams run through during ML model development to showcase the metadata management process in action: (1) profiling the existing data sources, (2) understanding the model performance against different fairness metrics, and (3) combining different explainability approaches to consider "what if" scenarios. For each step, we also illustrated how existing transparency artefacts can be utilised to communicate the findings to other stakeholders in the development process and broader audience. Figure \ref{fig:metadata} illustrates a generic workflow of recording and communicating these metadata information.

\subsubsection{Metadata of Indian News Dataset}

\begin{itemize}
    \item \textbf{Base metadata:} Although the dataset is an open dataset, stored in Huggingface, it misses key information about fairness, specifically, representativeness, and data collection methodologies.
    \item \textbf{Record representation:}  We utilised a pre-trained named-entity-recognition (NER) model (Stanza \cite{qi_stanza_2020}) to identify the the words that can relate to sensitive features (country, currency, etc.) to estimate overall representativeness of the dataset. (For example, some entities are \textit{RBI (Reserve Bank of India), NSE (National Stock Exchange), BSE (Bombay Stock Exchange), INR (Indian Rupee), Lakh (100,000), Aadhar (Identification Authority)}) The NER model identified 5622 samples containing organisation or country-related information, which could implicitly reveal country-specific details. 
\end{itemize}

\noindent \fcolorbox{main}{sub}{ 
\begin{minipage}{\dimexpr\linewidth-2\fboxsep-2\fboxrule} 
Although some datasets come with a thorough documentation and complete metadata, most datasets does not contain adequate information regarding the representation of sensitive characteristics. In this stage, the development team can map the obtained metadata to a selected dataset card format containing responsible/fair AI entities. Further, the team should record changes throughout the cleaning and preprocessing. Documenting the balance of represented groups to note any inherent biases is essential to prevent future potential discrimination cases.
\end{minipage}
}

\subsubsection{Metadata of FinBERT performance analysis}

\begin{itemize}
    \item \textbf{Base metadata:} The model card for FinBERT can be accessed via Huggingface, a repository for machine learning models and datasets. Additionally, since FinBERT is based on a BERT-base model, the team can refer to BERT's model card to gather further insights into the capabilities and limitations of the base model. Complementary information can also be obtained from FinBERT's research paper.
    \item \textbf{Model Description - Performance Metrics:} Compared to the original dataset performance evaluation of FinBERT, which was reported as 0.88 \cite{huang_span_2023}, the model's performance on the Indian News dataset is significantly lower. The model's three-category classification accuracy on the Indian News dataset is 0.57.
    \item \textbf{Model Description - Performance Metrics:} To evaluate the model's performance across different groups in the synthetic and real datasets we reported. We tested based on demographic parity, equalised odds, and equal opportunity notions, by using Fairlearn library \cite{fairlearn}. However, the model did not yield a significant different based on these fairness notions. (Note that, we recorded these information under performance metrics, as model card schema does not contain any subgroup metric recording entities.)
\end{itemize}

\noindent \fcolorbox{main}{sub}{ 
\begin{minipage}{\dimexpr\linewidth-2\fboxsep-2\fboxrule} 
After collecting these resources, the team must identify any gaps in the provided information and determine how to address the risks associated with these gaps. This process involves analyzing and documenting key metrics such as accuracy, precision, recall, F1-score, and confusion matrices, segmented across various demographic groups, financial sectors, regions, or types of financial instruments.
\end{minipage}
}

\subsubsection{Metadata of a token-level "what-if" experiment}
\hfill

\noindent
\fcolorbox{main}{white}{
\begin{minipage}{\dimexpr\linewidth-2\fboxsep-2\fboxrule} 
\textbf{Example sentence from dataset:} \textit{New car registrations collapsed by a 'precipitous' 97 percent last month. decline is in line with similar falls across Europe. many showrooms were closed for the coronavirus lockdown. around 1.68 million new cars will be registered in 2020. the lockdown was implemented nationwide on march 23. a strong new car market supports a healthy economy.}
\end{minipage}
}

We used the Learning Interpretability Tool (LIT) \cite{tenney_language_2020} to run and visualise the token-level importance to test an individual phrase, such as "across Europe", contributes positively to the \textit{Positive} sentiment tone in a sample, even though the true label was \textit{Negative}.

Despite the label being Negative, the model predicted \textit{Positive} with 100\% confidence. To investigate, we analyzed the impact of the word "Europe" by replacing it with other regions (e.g., Asia, Africa). The prediction remained \textit{Positive}, suggesting that "Europe" did not significantly influence the outcome according to LIME or IG. This points to a reliability issue rather than a fairness concern.

We identified two key factors affecting the model's decision: (1) Adding a strong positive statement can shift the prediction (e.g., moving the last sentence to the start dropped the Positive prediction score to 0.75), and (2) Large numbers can influence predictions (e.g., removing the sentence about '1.68 million new cars' changed the prediction to \textit{Negative}). We recorded this information in fairness metadata (under data - sample entity) and flag it in the risk register for the consideration of other teams.

\noindent \fcolorbox{main}{sub}{ 
\begin{minipage}{\dimexpr\linewidth-2\fboxsep-2\fboxrule} 
    Interpreting interpretability methods like LIME and Integrated Gradients (IG) is challenging due to the lack of standardised approaches \cite{garreau20a}.  However, these methods can provide valuable starting points for identifying words or phrases that significantly influence a model's decisions and uncovering patterns from the bottom up.
\end{minipage}
}

\subsubsection{Syncing Metadata Information}

During the fairness-related experiments (in this case: 1. NER-based representation analysis, 2. fairness across sub-groups, and 3. token-level explanations for selected sentences), the results are stored in a dedicated fairness log file. A separate log file is generated for each experiment, as illustrated in Figure \ref{fig:metadata}. This information can later be synced with data and model card metadata to ensure the latest fairness insights are included and shared with stakeholders, including the public. Our monitoring toolkit, available on GitHub, facilitates seamless integration and metadata synchronization throughout the ML development pipeline.

\section{Discussion}

Although we designed our approach with a multi-stakeholder focus, the implementation of argument-based assurance can present some challenges in an environment with multiple levels and cycles of processes. The assurance case creation still requires a level of understanding and expertise of the current opportunities and risks, and organizations must be prepared to address potential trade-offs between fairness and other objectives. Furthermore, the reliance on comprehensive data and documentation introduces the challenge of data sensitivity and the need to prevent information overload, while the evolving nature of fairness standards requires ongoing adaptation. Successfully integrating this approach requires careful consideration of context-specific requirements, dynamic environments, and the need for tools to support integration with existing systems. 

Despite these challenges, the proposed framework offers a significant step towards responsible AI development. It aligns well with existing risk management structures (e.g. NIST, EU AI Act, UK Assurance Framework, ISO 23894) by enabling feedback mechanisms through risk and compliance teams as well as business analysts. It can help streamline risk assessment while providing clear and transparent artefacts (e.g. assurance cases). By providing templates and utility functions, the monitoring flow supports standardised metadata usage, reducing the manual effort typically required in these processes. Instead of adding redundant layers, our approach enhances existing documentation practices. The technical adaptability and modularity of this approach can allow developers to adopt tailored versions for various organizational contexts and stages of maturity.

Justified trust in AI systems hinges on the effective communication of reliable evidence to users and stakeholders. In a commercial setting, our approach can support the distribution of shared responsibility for building this justified trust between IT teams (development and testing) and business teams (analysts and compliance). This collaboration can result in more effective risk management flows, where risk refers to potential technical (data, model, and interaction) problems that could adversely impact project success. To facilitate a well-organised risk governance process and improve knowledge sharing, we put our current effort into structuring the evidence collection process to align with existing risk management frameworks. For example, by integrating the RAID (Risks, Assumptions, Issues, Dependencies) methodology into metadata formatting, we are automating the transfer of risks into GitHub issues. This can enable teams to address risks systematically by providing evidence, such as experiment results, while strengthening the connection between risk identification and resolution.

\section{Conclusion and Future Work}

This paper introduces a two-step framework, as a keystone towards achieving justified trust in AI-enabled system development. It uses argument-based assurance, a structured approach to justifying AI system decisions, combined with dynamic evidence collection from various transparency artefacts like model cards and data cards. This dynamic process addresses the challenge of transferring algorithmic solutions between differing social contexts, portability trap, by making assumptions explicit and monitoring evidences in need. It supports development teams to build a proactive and systematic fairness assessment, reviewing, and monitoring by defining goals, gathering evidence, and continuously monitoring system performance. 

The effectiveness of this framework is demonstrated through an illustrative case study in finance, focusing on supporting fairness-related arguments. The use case motivation was exploring the implicit bias sources that can lead to more positive interpretation of data dominant regions, based on the Global North/South definition. We collected evidence from  three tasks, including (1) profiling the existing data sources, (2) understanding the model performance against different fairness metrics, and (3) combining different explainability approaches to consider "what if" scenarios. Throughout this use case, we demonstrated practical metadata recording suggestions for development teams. The metadata management library and demo use case is available on Github (\href{https://github.com/alan-turing-institute/fairness-monitoring}{alan-turing-institute/fairness-monitoring}).

\section*{Positionality Statement}
The authors belong to a publicly-funded research organisation and collaborate closely with academia, government bodies, and the private sector. This positioning enables the team to gather informal, real-world feedback throughout the design and development process. The research team is diverse in both professional backgrounds and nationalities, which organically allows us to consider multiple perspectives in framing and addressing fairness concerns.

\begin{acks}
This work was a part of \href{https://www.turing.ac.uk/research/research-projects/proactive-monitoring-ai-fairness}{Proactive Monitoring of AI Fairness} research, which is supported by Innovate UK [Project No: 10108523]. This work was also supported, in whole or in part, by the Gates Foundation [INV-057591]. Additional funding was provided by UKRI through the EPSRC grants: EP/V056883/1 (Framework for Responsible AI in Finance) and EP/R007195/1 (Academic Centre of Excellence in Cyber Security Research - University of Warwick)
\end{acks}

\bibliographystyle{ACM-Reference-Format}
\bibliography{bibliography}

\end{document}